\documentclass[12pt]{article}
\usepackage[a4paper,margin=2.5cm]{geometry}
\usepackage{amsmath,amssymb}
\usepackage{booktabs}
\usepackage{graphicx}
\usepackage{placeins}
\usepackage{caption}
\usepackage{xcolor}
\usepackage[colorlinks=true,allcolors=blue]{hyperref}

\graphicspath{{figure/}}
\providecommand{\ack}{\section*{Acknowledgments}}
\providecommand{\etal}{\textit{et al}}

\begin{document}

\begin{center}
{\LARGE Active learning at a graduate summer school:\\[2pt]
feasibility, student response, and the role of instructor training\par}
\vspace{1.2em}
{\large Nicholas Kamp$^1$, Louis Deslauriers$^{1,2}$, Carlos A. Arg\"uelles$^1$, Gonzalo Herrera$^{1,3}$ and William G. Thompson$^{1}$\par}
\vspace{0.8em}
{\small $^1$ {Department of Physics \& Laboratory for Particle Physics and Cosmology, Harvard University, Cambridge, MA 02138, USA\\
$^2$ Faculty of Arts and Sciences, Harvard University, Cambridge, MA 02138, USA\\
$^3$ Kavli Institute for Astrophysics and Space Research, Massachusetts Institute of Technology, Cambridge, MA 02139, USA\par}}
\vspace{0.6em}
{\small E-mail: \href{mailto:louis@physics.harvard.edu}{louis@physics.harvard.edu}\par}
\end{center}
\vspace{0.5em}

\begin{abstract}
\noindent Active learning is becoming standard practice in undergraduate STEM instruction in North America and is slowly being adopted in graduate courses. Graduate summer schools, one- to two-week programs in which students from many institutions gather for intensive, specialized instruction in their own field, are common in physics; they are taught by researchers chosen for their expertise in the subject, and their pedagogy remains almost entirely traditional lecturing. In this paper we report a study at a one-week graduate school in neutrino and astroparticle physics attended by an international cohort of 38 students from 8 countries. Six lecturers with limited or no prior experience of teaching with active learning introduced Peer Instruction-style active learning into their lectures after approximately four hours of preparation led by a physics-education specialist experienced in implementing active learning. Each of the three main lecture streams was taught traditionally on one day and in the active format on the other. Students were surveyed after each of the first two days and at the end of the week; all six instructors were surveyed as well. In every active lecture, spanning three instructors and three topics, 82–100$\%$ of respondents reported that the interactive segments helped them stay engaged and check their understanding, and in every stream the difficulty was rated as better matched on the active day. At the end of the week, 33 of 36 wanted such teaching in future graduate courses, and not one respondent disagreed. All six instructors judged the approach productive and intend to use it again; only two found implementation straightforward. With brief expert-led preparation, lecturers new to active learning implemented it successfully for a heterogeneous audience, and both students and instructors want more. The school was also the first time any of the six instructors implemented structured active learning: a summer school can develop its lecturers as well as its students.
\end{abstract}

\vspace{0.5em}

\vspace{1.2em}

\section{Introduction}\label{sec:intro}

Active learning, in which class time is spent reasoning, discussing, and receiving feedback rather than listening to exposition, improves learning in undergraduate STEM courses~\cite{Hake1998,CrouchMazur2001,DeslauriersWieman2011,Freeman2014} and is becoming standard practice in North America, although lecturing remains the most common observed mode~\cite{Stains2018}. At the graduate level, published implementations are fewer. Patton taught a first-year graduate electrodynamics course as structured group inquiry in the 1990s~\cite{Patton1996}. Carr and McKagan, in a three-year study of a graduate quantum mechanics course, found that graduate students respond well to research-based techniques developed for introductory courses~\cite{CarrMcKagan2009}; Porter and Heckler measured pre-to-post gains from guided group work accompanying a graduate quantum mechanics course~\cite{PorterHeckler2020}; and Lepage converted a semester-length graduate quantum field theory course to a reading-plus-worksheet active format, refined over four consecutive iterations by a single instructor~\cite{Lepage}. Incoming graduate students carry many of the conceptual difficulties documented in undergraduates~\cite{Singh2008}, and graduate students themselves report wanting more active learning than they receive~\cite{Gonsar2021}.

Summer schools are a significant part of many graduate students' training. In these one- to two-week programs, students from many institutions, often meeting for the first time, gather for intensive instruction in a specialized area of their field. Their teaching has remained almost entirely traditional lecturing, and calls for active teaching at such schools note that their effectiveness has not been studied~\cite{Moldwin2023}. The one published account we have found is the school of L\'opez and Gross, who restructured a two-week graduate space-weather school around active engagement (Peer Instruction, daily formative evaluation, jigsaw groups, and concept mapping) with a dedicated education team, and reported consistently favorable participant ratings~\cite{LopezGross2008,GrossEOS2009}. A typical school has no such team: its lecturers, researchers chosen for their expertise in the subject, each teach once or a few times, to an audience whose preparation in the subfield varies widely. Whether these lecturers can implement Peer Instruction-style active learning after brief preparation, and how such an audience and the lecturers themselves respond, has not been reported.

We report a study at a one-week graduate school in neutrino and astroparticle physics attended by 38 students from 8 countries, heterogeneous in subfield preparation. Ten lecturers were invited, all early-career researchers in neutrino astrophysics without a permanent position. Six of them took part in the study: the three who taught the core lecture streams and three who taught specialized sessions. All six had limited or no prior experience of teaching with active learning, and introduced Peer Instruction-style active learning into their lectures after approximately four hours of preparation led by a physics-education specialist experienced in implementing active learning. Each of the three core lecture streams was taught traditionally on one day and in Peer Instruction style on the other; students were surveyed after each of those days and again at the end of the week, and the six instructors were surveyed on their experience. All outcomes are self-reported perceptions. Because none of the six lecturers had implemented structured active learning before, the school was also their first supported experience of teaching this way. The results therefore bear on summer schools as a development opportunity for early-career instructors, and not only as instruction for students.

\section{Setting and implementation}\label{sec:setting}

\subsection{The school}
The study took place at the Cosmic Neutrino Practicum, a one-week graduate summer school in neutrino and astroparticle physics at Harvard University. The core program was a three-day sequence of morning lectures, three 75-minute lectures on each of Monday, Tuesday, and Wednesday, with specialized afternoon sessions and further sessions later in the week; we refer to the morning lectures as the core lectures. A total of 38 graduate students attended; 38 responded to the Day-1 survey, 36 to the Day-2 survey, and 36 to the end-of-week retrospective. Responses were anonymous and not linked across days, so we do not know which students answered on both days. All surveys were optional. As the study used anonymous, voluntary feedback surveys collected for the school's own program evaluation, it was exempt from Institutional Review Board oversight. Instructor responses are reported without names.

\subsection{Format rotation}
The core lectures were organized as three streams, numbered by their fixed order in the morning schedule: Stream~1 began at 9:00 and Stream~3 occupied the final late-morning slot. Each stream had its own topic and its own instructor, with Instructor~$n$ teaching Stream~$n$ on all three days. Across Monday and Tuesday each stream kept its topic and instructor while the format changed (table~\ref{tab:design}). All participants experienced the same schedule; there was no randomization or counterbalancing. Within-stream day-to-day comparisons are therefore confounded with day-specific content and ordering, and comparisons across streams are additionally confounded with instructor, topic, and time of day. The Wednesday core lectures were laboratory-style and inherently interactive, providing no traditional contrast, and are excluded; afternoon and later-week sessions were specialized and outside the design.

\begin{table}[h]
\centering
\caption{Format rotation of the core lectures across the two comparison days. Instructor $n$ taught Stream $n$ on all three days; between Monday and Tuesday the format changed, and so did the day's content.}
\label{tab:design}
\begin{tabular}{lllll}
\toprule
Stream & Instructor & Topic & Day 1 & Day 2  \\
\midrule
1 & Instructor 1 & Neutrino Telescopes & Traditional & Active \\
2 & Instructor 2 & Neutrino Interactions & Traditional & Active \\
3 & Instructor 3 & Reconstruction/Machine Learning & Active & Traditional \\
\bottomrule
\end{tabular}
\end{table}

\subsection{The active-learning episodes}
We used Peer Instruction~\cite{Mazur1997,CrouchMazur2001}, chosen because it can be inserted into an existing lecture without redesigning it, needs no equipment, and is simple enough to rehearse in an afternoon~\cite{LasryMazurWatkins2008}. A conceptual question, usually open-ended and occasionally multiple-choice, was displayed on a slide; students thought about it individually, then discussed it with their neighbors while the instructor circulated among the groups, and the instructor then led a whole-class debrief. We refer to each such question-discussion-debrief cycle as an \emph{active-learning episode}; three of the questions used, with what the discussion produced, are given in appendix~\ref{app:episode}. Two features depart from Peer Instruction as usually described, as implementations commonly do~\cite{TurpenFinkelstein2009}: most questions were open-ended rather than multiple-choice, and there were no clickers, which has been found not to affect learning~\cite{Lasry2008}; where votes were taken they were by show of hands or, on Day~2, a web polling tool. In the Day-1 active lecture, six episodes occurred at roughly ten-minute intervals, averaging about five minutes each; counting the full episode (individual thinking, discussion, and debrief), about 40\% of the lecture was active.

\subsection{Instructor preparation}
The six instructors in the study received the same preparation: (i) a video-call discussion ($\sim$1.5\,h) explaining the rationale and mechanics of the episodes, and (ii) an in-person session ($\sim$2.5\,h) the Sunday before the school in which each instructor rehearsed their active segments while the others role-played students, with real-time feedback from the physics-education specialist. All six instructors attended both components. The three rotation instructors were asked only to insert conceptual questions with short peer discussions into their active-day lecture, not to redesign it; the three instructors of specialized sessions received the same preparation but were not asked to use it. The roughly four contact hours exclude instructors' independent preparation time, which was not recorded.


\section{Data and analysis}\label{sec:data}

\subsection{Instruments}
The Day-1 and Day-2 surveys used five-point Likert items. For every lecture, students rated whether the ideas were genuinely interesting and whether the difficulty felt about right. For the active lectures they also rated whether the interactive parts helped them stay engaged and helped them check their understanding, and, once per day, whether they wanted to continue seeing interactive elements. One open item asked what was valuable. On Day~2, two additional items asked whether students wished Stream~3, now taught traditionally, had included interactive elements.

The end-of-week retrospective had four Likert items on the week's teaching and four on general satisfaction with the school (given in appendix~\ref{app:full} as context). One item asked whether the amount of interactive time was about right. Students then selected every element they had found valuable, and the single most valuable one. Two free-response items asked whether their view of the elements had changed over the week and what could be improved.

The instructor survey covered teaching background, prior use of active learning, ratings of each preparation component, per-lecture reports of deliberate active elements and the fraction of class time they occupied, the experience of implementing them, and free text. One item asked whether an inexperienced instructor could implement the elements effectively with no training at all; its four response options appear verbatim with the results (table~\ref{tab:instructors}). The full instruments are provided as supplementary material.

\subsection{Analysis}
We report counts and percentages. With 36--38 respondents, a 95\% confidence interval on a proportion near 80\% is about $\pm13$ percentage points, and the only comparisons available across days are confounded by content and ordering, so we do not report significance tests. Where we report how many free-text responses mention a theme, a response was counted only if it clearly referred to that theme. Full Likert distributions are given in appendix~\ref{app:full}. Aggregate Day-1 results were shared with the organizer during the week; the instructors did not see them until after the school.

\section{Results}\label{sec:results}

\subsection{Implementation}
All three active lectures ran their episodes within the allotted time, with no cancellations or technical failures. The instructors estimated the fraction of class time students spent actively working at 10--25$\%$ for the Day-1 active lecture and 25--50$\%$ for each of the two Day-2 active lectures; for the Day-1 lecture, counting full episodes, our own observation gives about 40$\%$. In that lecture, roughly 37 of 38 students were visibly on task in each episode.

\subsection{Student ratings across the rotation}
\begin{figure}[h]
\centering
\includegraphics[width=0.85\textwidth]{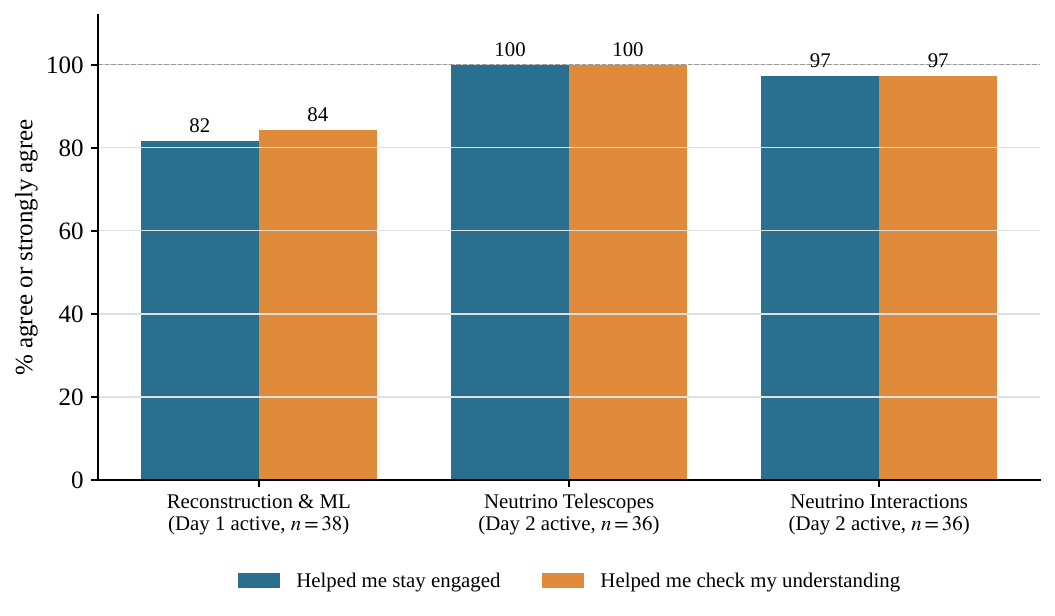}
\caption{Ratings of the interactive elements in the three active lectures, spanning three instructors and three topics. Bars show the percentage of respondents who agreed or strongly agreed ($n=38$ Day 1; $n=36$ Day 2). Full distributions are given in appendix~\ref{app:full}.}
\label{fig:ratings}
\end{figure}

Ratings of the interactive elements were favorable in all three active lectures (figure~\ref{fig:ratings}; counts in table~\ref{tab:interest_diff} and appendix~\ref{app:full}). In the Day-1 active lecture (Reconstruction and Machine Learning), 31/38 respondents (81.6$\%$) agreed or strongly agreed (hereafter ``agreed'') that the interactive parts helped them stay engaged and 32/38 (84.2$\%$) that they helped them check their understanding, with 30/38 (78.9$\%$) wanting more such elements. In the two Day-2 active lectures, the corresponding figures were 36/36 and 36/36 (Neutrino Telescopes) and 35/36 and 35/36 (Neutrino Interactions), with 33/36 (91.7$\%$) endorsing a single joint item on continuing interactive elements. The three lectures were rated by the same cohort, so they are not independent replications, but the favorable response was not specific to one lecturer or subject.

Interest and difficulty behaved differently (table~\ref{tab:interest_diff}). Interest was high for all Day-1 lectures and for the two Day-2 active lectures (33--37 of the respondents per item), but fell to 17/36 (47.2\%) for the traditional Day-2 delivery of Reconstruction and Machine Learning. The difficulty ratings show two patterns. First, the Reconstruction and Machine Learning stream had the lowest ``about right'' rating on both days (16/38, then 6/36), and open comments repeatedly described it as difficult. Second, within every stream the ``about right'' percentage was higher on that stream's active day (81.6$\to$88.9, 63.2$\to$91.7, 16.7$\to$42.1, ordering each stream traditional$\to$active). Because format changed together with day-specific content and ordering, neither pattern supports a causal reading in either direction.

\begin{table}[h]
\centering
\caption{Respondent endorsement of interest and difficulty-calibration questions, as count (\%) agreeing or strongly agreeing, by stream and day. Bold marks the active delivery. $n=38$ (Day 1), $n=36$ (Day 2).}
\label{tab:interest_diff}
\begin{tabular}{llcccc}
\toprule
 & & \multicolumn{2}{c}{Interest} & \multicolumn{2}{c}{Difficulty ``about right''} \\
\cmidrule(lr){3-4}\cmidrule(lr){5-6}
Stream & Topic & Day 1 & Day 2 & Day 1 & Day 2 \\
\midrule
1 & Neutrino Telescopes & 37 (97.4) & \textbf{34 (94.4)} & 31 (81.6) & \textbf{32 (88.9)} \\
2 & Neutrino Interactions & 34 (89.5) & \textbf{35 (97.2)} & 24 (63.2) & \textbf{33 (91.7)} \\
3 & Reconstruction and ML & \textbf{33 (86.8)} & 17 (47.2) & \textbf{16 (42.1)} & 6 (16.7) \\
\bottomrule
\end{tabular}
\end{table}

Students did not reserve their support for interactivity for the easier material. When Reconstruction and Machine Learning, the least well-calibrated stream, was taught traditionally on Day 2, 24/36 respondents (66.7\%) wished it had included interactive elements to keep them engaged and 25/36 (69.4\%) to help them check their understanding. These items differ in wording from the Day-1 and Day-2 items on continuing interactive elements, so the percentages should not be ranked against each other.

\subsection{Open comments during the week}
Day-1 free responses (35 of 38 respondents) and Day-2 free responses (25 of 36) were coded for recurring themes. Peer discussion was the component most often named as valuable (mentioned in 19 of the 35 Day-1 responses), typically for checking one's understanding against others and surfacing shared confusion (``check misinterpretations of the content''; ``against other students rather than just wait for an answer from the instructor''). Critical comments concerned facilitation rather than the format itself: confusingly posed questions, insufficient scaffolding, episodes running long, a request for reference material kept visible, and, in two Day-1 responses, an instructor's discouraging tone when receiving answers. One Day-2 respondent preferred the conceptual questions to the previous day's calculation-oriented ones; another noted that the traditional Day-2 lecture ``was also very good without them.''

\subsection{End-of-week ratings}
\begin{figure}[h]
\centering
\includegraphics[width=0.98\textwidth]{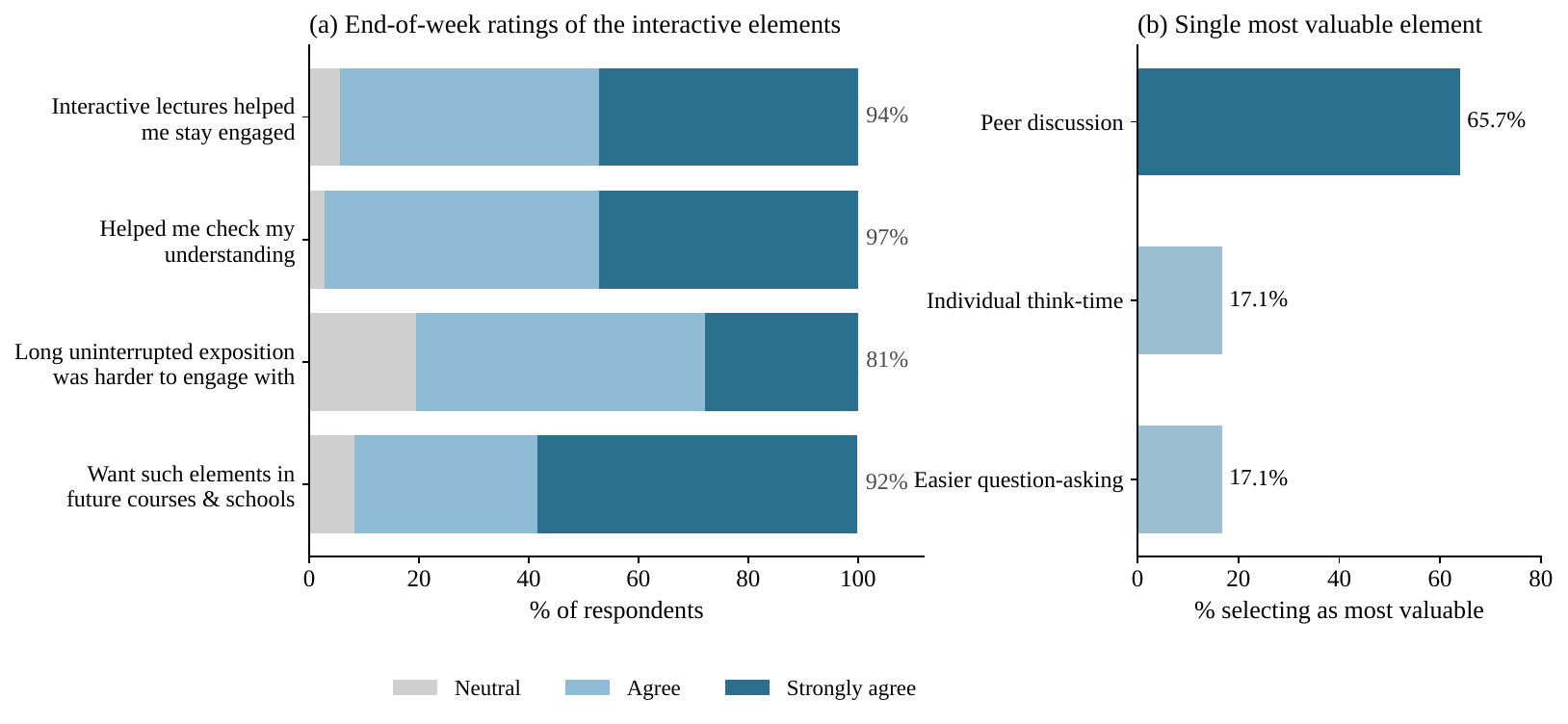}
\caption{End-of-week retrospective ($n=36$). (a) Ratings of the four teaching items; stacked bars show neutral, agree, and strongly agree; no respondent selected disagree or strongly disagree. (b) Element selected as single most valuable; one respondent selected two options and is excluded from the bars as a percentage of the 35 respondents who selected one option.}
\label{fig:retro}
\end{figure}

At the end of the week, 34/36 (94.4\%) agreed that the interactive lectures had helped them stay engaged, 35/36 (97.2\%) that the interactive segments had helped them check their understanding, 29/36 (80.6\%) that long stretches of uninterrupted exposition had been harder to engage with, and 33/36 (91.7\%) that they would like such elements in future graduate courses and schools (figure~\ref{fig:retro}(a)). No respondent disagreed with any of the four. These items were all worded in the direction favorable to interactivity, and the survey was anonymous and unlinked, so it cannot separate a durable preference from novelty or from a wish to please. The favorable response of the first two days had not faded by the last.

On the amount of interactive time over the week, 29/36 judged it about right, 5 too little, and 2 too much. Asked which elements they had found valuable, 32/36 checked peer discussion, 30/36 individual think-time, and 18/36 easier question-asking; asked for the single most valuable, 35 respondents selected one option and one selected two, and of the 35, 23 (65.7\%) chose peer discussion and 6 (17.1\%) each chose the other two (figure~\ref{fig:retro}(b)).

Asked whether their view of the interactive elements had changed over the week (35 responding), 9 said they had warmed to them, 25 reported no change, most adding that they had liked them from the start, and 1 had cooled, citing occasionally unclear questions. One respondent wrote of the lectures without interactive elements, ``when they were taken out I definitely missed them.''

Suggestions for improvement concentrated on two points of facilitation: reference material kept visible during discussion (15 of 34 substantive responses, 13 naming reference slides explicitly) and clearer questions (9 of 34), with scattered requests for more scaffolding, better pacing, and more circulating facilitators (``there was one instructor walking around to help many groups, so we often spent a long time waiting to ask a question'').

\subsection{Instructor experience and views of the preparation}
All six instructors in the study completed the survey: the three who taught the rotation streams (I1--I3 in table~\ref{tab:instructors}) and three who taught specialized afternoon sessions (S1--S3). Their self-reports were given to the author who trained them (LD), and two authors (NK and GH) are among the six respondents.

The instructors had varied teaching backgrounds (teaching-assistant work, exercise classes, laboratory instruction, tutoring); three of six, including two of the three rotation instructors, had never used active-learning elements in their own teaching, and the other three had used them only informally. One had completed a 40-hour online course on evidence-based teaching. None had implemented structured active-learning episodes of the kind used here.

All six agreed the combined preparation was important to implementing the elements well (three strongly), and all six felt adequately prepared. On whether an inexperienced instructor could implement the elements effectively with no training, the verbatim distribution was: ``No, some minimum training is necessary'' 1/6; ``Probably not'' 3/6; ``Probably, but training helps'' 2/6; 
``Yes, no training needed'' 0/6. Because every instructor received the same preparation, the study cannot estimate its causal effect or a minimum effective amount. Two instructors suggested changes: one would weight the preparation further toward the in-person rehearsal, the other asked for \emph{more} coverage of the pedagogical rationale.

All six perceived the episodes as productive for their students, and all six would use them again (three strongly on each item); yet only two agreed implementation was straightforward, with two neutral and two disagreeing. Named difficulties spread across several areas: authoring conceptual questions at the right level (2/6; one reported this eased once the questions were aimed at common misconceptions), controlling episode length while circulating (1), the tension between interactive time and content coverage (1), judging when to intervene in discussions (1), discomfort soliciting whole-class participation (1), and some students preferring to think alone (1). One rotation instructor reported the active portion was ``much easier the second time''; another, who had taught exercise classes for several years but had never used active learning, wrote that he ``was surprised how engaged the students were in the active learning discussions.''

\begin{table}[h]
\centering\small
\caption{Instructor census ($n=6$). I$n$ denotes the instructor of Stream~$n$ (table~\ref{tab:design}); S1--S3 taught specialized afternoon sessions. Likert: SA strongly agree, A agree, N neutral, D disagree. The final column gives the verbatim option selected on whether an inexperienced instructor could implement the elements effectively with no training.}
\label{tab:instructors}
{\footnotesize
\begin{tabular}{lcccccp{3.4cm}}
\toprule
 & Prior active & Prep.\ & Straightforward & Perceived & Use & ``No training''\\
Instructor & learning & important & to implement & productive & again & item (verbatim) \\
\midrule
I1 & Informal & SA & A & SA & SA & Probably not \\
I2 & Never & SA & N & SA & SA & Probably not \\
I3 & Never & A & D & A & A & No, some minimum training is necessary \\
S1 & Informal & SA & N & A & SA & Probably not \\
S2 & Informal & A & A & SA & A & Probably, but training helps \\
S3 & Never & A & D & A & A & Probably, but training helps \\
\bottomrule
\end{tabular}}
\\[2pt]
\end{table}

\subsection{Adoption beyond the assigned lectures}
Instructors of specialized sessions, who received the same preparation and attended the core lectures but were not asked to use active elements, reported deliberately including them in some of their own lectures (self-reported fractions of class time from under 10\% to 10--25\%), consistent with the authors' informal observation over the week.

\section{Discussion}\label{sec:discussion}

After roughly four hours of common preparation, early-career lecturers with limited or no prior experience of structured active learning delivered brief conceptual-question and peer-discussion episodes in a graduate summer school. Respondents rated the episodes favorably in each of three lecture streams and at the end of the week. The instructors found them productive and intend to use them again, though most did not find them easy. Each rotation instructor taught both formats, so within each stream the traditional and active days had the same instructor and topic. On that comparison the active day was rated better calibrated in all three streams, and interest was similar in two and much higher on the active day in the third. L\'opez and Gross~\cite{LopezGross2008} rebuilt a two-week program around active learning with a dedicated education team. Here, active-learning episodes inserted into otherwise conventional lectures after a few hours of shared practice also worked and were well received. Lepage's graduate quantum field theory course~\cite{Lepage} ran for a semester with one instructor. This was one week with several instructors.

The choice of pedagogical method is, in our view, the main reason this worked. Peer Instruction-style episodes ask a lecturer to do one new thing, pose a question and let students discuss it before the debrief, and the version used here, without clickers, is the simplest form of the method. More demanding forms of active learning, such as the worksheet-based deliberate-practice format in which the whole class period is restructured around student work and feedback~\cite{DeslauriersWieman2011}, have been shown to produce large learning gains, but a lecturer new to them needs an expert mentor alongside for a semester~\cite{McCartyDeslauriers2020}. We would not recommend attempting these formats with lecturers teaching this way for the first time. For a summer school, active-learning episodes within otherwise conventional lectures are a sensible first step.

The study does not show learning gains: all outcomes are perceptions, and perceived learning under active instruction can diverge from measured learning, and can run opposite to it~\cite{Deslauriers2019}, so the present favorable ratings must not be read as evidence of effectiveness. It does not show an advantage over traditional lecturing: format changed together with day, content, and ordering, and interest and calibration patterns admit no causal reading. It does not show that the preparation was necessary: all instructors received it, so its causal contribution and minimum amount are untested. All six valued it and none thought it unnecessary. And it does not identify peer discussion as a mechanism in the experimental sense of Smith \etal~\cite{Smith2009}. It identifies peer discussion as the component respondents most often selected as valuable. The lecturers were early-career researchers recruited by the organizers, not invited senior lecturers, and the survey items were written for this study and have not been validated.

The free-text responses from students and instructors agree on where the difficulty lies: not in posing a question but in running the discussion that follows. Students named unclear questions, missing reference material during discussion, episode length, and, in two cases, a discouraging tone from the instructor when answers were offered. Instructors named question authoring, timing, and knowing when to intervene. This matches the observation that implementations of nominally identical active-learning structures vary widely in practice~\cite{TurpenFinkelstein2009}. Inserting a question is the easy part. How it is posed, supported, and debriefed determines how students respond. One end-of-week comment put it this way: the lectures ``became much less `conference proceedings to peers' and much more `helping students learn'.''

The other implication is for the instructors. Six early-career researchers here received their first structured experience of active-learning instruction under supportive conditions: a rehearsal with feedback, at least one live implementation, and immediate, visible student response. One rotation instructor reported the second attempt was much easier than the first, the specialized-session instructors adopted elements voluntarily, and all six intend to use the methods again. Intention is not persistence: in an early national survey about a third of physics faculty who had tried research-based methods had stopped using them~\cite{Henderson2012}. A later survey found about one in twenty~\cite{Dancy2024}, and among faculty whose first implementation was supported by a science-education specialist almost none stopped~\cite{Wieman2013}. The preparation here was a brief form of that support.
A summer school is a low-stakes place for early-career physicists to teach this way for the first time, and the school gets that for the cost of the preparation. Whether that experience transfers into the instructors' subsequent teaching was not measured here and is worth following up.

To try this, have the lecturers practice their own active-learning episodes in front of each other, with the others playing the role of students. Five of the six instructors rated that rehearsal valuable. Write conceptual questions that target the common difficulties and misconceptions of the subfield~\cite{Mazur1997,Beatty2006}, and leave reference material on the screen during the discussion, the most frequent request in the end-of-week survey. Keep each active-learning episode to a few minutes, at roughly ten-minute intervals, as in the Day-1 lecture here and in Peer Instruction as usually practiced~\cite{CrouchMazur2001}. Have enough people circulating that groups do not wait long for help~\cite{Lepage}, as students here asked for. Receive student answers in a supportive tone. How instructors handle answers is one of the ways implementations differ~\cite{TurpenFinkelstein2009}, and students notice: two Day-1 comments were about an instructor's tone.

Three things make a graduate summer school different from a graduate course. The students typically come from many departments and institutions, so their preparation and their prior exposure to active learning can be expected to vary more than in a course. There is no time to adjust: Lepage notes that some of his students took time to get used to group work~\cite{Lepage}, and he had a semester, whereas here the episodes had to work in the first hour. And the lecturers change from session to session, each teaching once, typically with no training. Results from undergraduate and graduate courses therefore cannot be assumed to hold here. The present study shows that untrained lecturers can be prepared in a few hours, and that a mixed cohort with no time to adjust responded favorably all the same. Learning should next be measured against a comparison condition. The preparation should be varied, to find which part of it matters and how little suffices. Group formation deserves attention in a cohort this mixed: whether self-selected groups serve students as well as groups formed with the instructor's help after a brief background survey is unknown. The episodes have been tried in one form only, and their length, frequency and question type are all open. None of this is particular to physics. Intensive, team-taught short courses with mixed participants exist across the sciences, as summer schools, training workshops and conference tutorials, and what is learned at one applies to the others.

\section{Conclusion}\label{sec:conclusion}
Students at an international graduate physics summer school responded well to Peer Instruction-style active learning. Each of the three core instructors taught one day traditionally and one day in the active format. Students rated the episodes favorably in every active lecture, across three instructors and three topics. In every stream they also rated the difficulty as better matched to them on the active day. At the end of the week 33 of 36 wanted such teaching in future graduate courses. The lecturers who delivered it had limited or no prior experience of active learning and about four hours of preparation with a physics-education specialist. All six found the episodes productive and intend to use them again, although only two found implementation straightforward. A graduate summer school can therefore develop its lecturers as well as its students. Active learning is becoming standard practice in undergraduate teaching, while the graduate summer school, with its mixed cohort, its single week and its changing lecturers, has remained a lecture format. This case shows that a few hours of preparation with a specialist are enough to begin changing that, and offers a model other schools can adopt.

\ack
We thank the instructors and participants of the \textit{Cosmic Neutrino Practicum}, and Molly Neylan for organizational support. 
We thank the Heising-Simons Foundation for supporting this summer school and this research through Grant \#2025-6482.
N.K. is supported by the National Science Foundation (NSF, CAREER Grant \#2239795) and the David \& Lucile Packard Foundation. 
C.A.A. are supported by the Faculty of Arts and Sciences of Harvard University, the John Templeton Foundation (Grant \#63651), the Research Corporation for Science Advancement, and the David \& Lucile Packard Foundation. 
N.K. and C.A.A. are supported by the National Science Foundation under Cooperative Agreement PHY-2019786 (The NSF AI Institute for Artificial Intelligence and Fundamental Interactions, http://iaifi.org/).
W.G.T. is supported by the John Templeton Foundation (Grant \#63651).
The work of G.H. was supported by the Neutrino Theory Network Fellowship with contract number 726844. Claude (Anthropic) was used to refine the text and in tabulating the survey data; all analyses and conclusions are the work of the authors.

\section*{Author contributions}
NK, CA, GH and WT organized the summer school.
NK taught one of the core lecture streams.
GH and WT each taught one of the specialized afternoon sessions.
LD provided the instructor preparation, observed the core lectures, and designed the surveys.
NK and LD analyzed the data.
LD wrote the first draft; all authors revised the manuscript.

\section*{Data availability}
The survey instruments are provided as supplementary material, and the full response distributions are given in the appendix. The individual responses, which include free-text comments and a six-person instructor census, are not shared publicly to preserve anonymity, but are available upon reasonable request.


\appendix
\section{Full response distributions}\label{app:full}
Counts of respondents in each category; $n=38$ on Day 1 and $n=36$ on Day 2 and in the retrospective. SD strongly disagree; D disagree; N neutral; A agree; SA strongly agree.

\begin{table}[h]
\centering\small
\caption{Day-1 and Day-2 survey items.}
\label{tab:fulldaily}
{\footnotesize
\begin{tabular}{llccccc}
\toprule
Item & Stream (day) & SD & D & N & A & SA \\
\midrule
Ideas genuinely interesting & 1 (Day 1) & 0 & 0 & 1 & 17 & 20 \\
 & 2 (Day 1) & 0 & 1 & 3 & 16 & 18 \\
 & 3 (Day 1, active) & 0 & 1 & 4 & 19 & 14 \\
 & 1 (Day 2, active) & 0 & 0 & 2 & 12 & 22 \\
 & 2 (Day 2, active) & 0 & 0 & 1 & 9 & 26 \\
 & 3 (Day 2) & 1 & 5 & 13 & 10 & 7 \\
\midrule
Difficulty about right & 1 (Day 1) & 0 & 0 & 7 & 15 & 16 \\
 & 2 (Day 1) & 1 & 4 & 9 & 15 & 9 \\
 & 3 (Day 1, active) & 3 & 10 & 9 & 15 & 1 \\
 & 1 (Day 2, active) & 0 & 1 & 3 & 13 & 19 \\
 & 2 (Day 2, active) & 1 & 0 & 2 & 13 & 20 \\
 & 3 (Day 2) & 6 & 13 & 11 & 6 & 0 \\
\midrule
Interactive parts helped me stay engaged & 3 (Day 1) & 0 & 2 & 5 & 24 & 7 \\
 & 1 (Day 2) & 0 & 0 & 0 & 15 & 21 \\
 & 2 (Day 2) & 0 & 0 & 1 & 12 & 23 \\
\midrule
Interactive parts helped me check understanding & 3 (Day 1) & 0 & 3 & 3 & 24 & 8 \\
 & 1 (Day 2) & 0 & 0 & 0 & 15 & 21 \\
 & 2 (Day 2) & 0 & 0 & 1 & 11 & 24 \\
\midrule
Would like to continue seeing interactive elements & (Day 1) & 0 & 1 & 7 & 17 & 13 \\
 & (Day 2) & 0 & 0 & 3 & 13 & 20 \\
\midrule
Wish Stream 3 had more interactive elements (engagement) & (Day 2) & 1 & 0 & 11 & 12 & 12 \\
Wish Stream 3 had more interactive elements (understanding) & (Day 2) & 0 & 0 & 11 & 13 & 12 \\
\bottomrule
\end{tabular}}
\end{table}

\begin{table}[h]
\centering\small
\caption{End-of-week retrospective. No respondent selected D or SD on any item. The four school-satisfaction items are given as context and are not part of the study's claims.}
\label{tab:retro}
\begin{tabular}{lccc}
\toprule
Item & N & A & SA \\
\midrule
\multicolumn{4}{l}{\emph{Teaching items}}\\
Interactive lectures helped me stay engaged & 2 & 17 & 17 \\
Interactive segments helped me check my understanding & 1 & 18 & 17 \\
Long uninterrupted exposition was harder to engage with & 7 & 19 & 10 \\
Want such elements in future courses and schools & 3 & 12 & 21 \\
\midrule
\multicolumn{4}{l}{\emph{School satisfaction (context)}}\\
Met my expectations & 1 & 15 & 20 \\
I feel like I learned a lot this week & 2 & 17 & 17 \\
Level well matched to my background & 9 & 21 & 6 \\
Would encourage a colleague to attend & 0 & 12 & 24 \\
\bottomrule
\end{tabular}
\end{table}

\FloatBarrier
\section{Active-learning episodes}\label{app:episode}
One question from each core stream, with what the discussion and the debrief produced.

\subsection*{Stream 2, Neutrino Interactions (Day 2)}
The question came after the charged-current and neutral-current channels had been introduced and the two event topologies, muon tracks and cascades, had been shown.

\medskip
\noindent\fbox{\begin{minipage}{0.95\textwidth}\small
\textbf{Question, on the slide.} Between muons (tracks) and electrons (cascades) in a neutrino telescope, which has the better energy resolution? Which has the better angular resolution?

\medskip
\textbf{Individual thinking, about one minute.} Students were asked to commit to both answers before talking.

\medskip
\textbf{Peer discussion, about five minutes.} The instructor circulated. The most common starting position was that tracks are better on both counts, because a track is the cleaner signature. Groups that moved off it usually had one member who pointed out that a through-going muon leaves the detector, so most of its energy is never measured.

\medskip
\textbf{Debrief, about two minutes.} A show of hands, then the two quantities in turn. A track gives a long lever arm, so the direction is well measured, below a degree at high energy, but the energy is only a lower bound. A cascade deposits its energy within tens of metres, so the energy is well measured, but the direction only to several degrees in ice. Neither topology is better at everything, which is why analyses select by topology according to the question being asked.
\end{minipage}}

\subsection*{Stream 1, Neutrino Telescopes (Day 2)}
At TeV energy scales, which statement best explains why the prompt neutrino flux is expected to eventually dominate over the conventional neutrino flux?
\begin{enumerate}
\item[(a)] Prompt neutrinos are produced closer to the surface of the atmosphere, so they lose less energy before detection.
\item[(b)] Charm hadrons decay before interacting, even at very high energies, so their neutrino yield remains energy-independent, unlike pions and kaons, which increasingly interact before decaying as energy rises.
\item[(c)] Prompt neutrinos carry a harder spectrum because charm quarks are heavier and are produced with more momentum than light mesons.
\item[(d)] At high energies, the atmosphere becomes optically thick to pions and kaons, which are absorbed before they can decay.
\end{enumerate}
The discussion turned on the competition between decay and interaction for each parent hadron; the debrief established that (b) is the reason, and that (c) names the right observation, a harder spectrum, with the wrong mechanism.

\subsection*{Stream 3, Reconstruction and Machine Learning (Day 1)}
What are the advantages, and the costs, of relying on the likelihood function for event reconstruction? The debrief collected the two sides: a likelihood fit models the detector response explicitly and returns uncertainties, but it is only as good as that model, it is expensive to evaluate, and it can get stuck in a local optimum; learned reconstructions trade some of that transparency for speed.


\begin{thebibliography}{99}


\bibitem{Hake1998} Hake R R 1998 Interactive-engagement versus traditional methods: a six-thousand-student survey of mechanics test data for introductory physics courses {\it Am. J. Phys.} {\bf 66} 64--74

\bibitem{CrouchMazur2001} Crouch C H and Mazur E 2001 Peer instruction: ten years of experience and results {\it Am. J. Phys.} {\bf 69} 970--7

\bibitem{DeslauriersWieman2011} Deslauriers L, Schelew E and Wieman C 2011 Improved learning in a large-enrollment physics class {\it Science} {\bf 332} 862--4

\bibitem{Freeman2014} Freeman S \etal\ 2014 Active learning increases student performance in science, engineering, and mathematics {\it Proc. Natl Acad. Sci. USA} {\bf 111} 8410--5

\bibitem{Stains2018} Stains M \etal\ 2018 Anatomy of STEM teaching in North American universities {\it Science} {\bf 359} 1468--70

\bibitem{Patton1996} Patton B R 1996 Group learning-based approach to the graduate electrodynamics course: Jackson by inquiry {\it APS Forum on Education Newsletter} (August) 25--6

\bibitem{CarrMcKagan2009} Carr L D and McKagan S B 2009 Graduate quantum mechanics reform {\it Am. J. Phys.} {\bf 77} 308--19

\bibitem{PorterHeckler2020} Porter C D and Heckler A F 2020 Effectiveness of guided group work in graduate level quantum mechanics {\it Phys. Rev. Phys. Educ. Res.} {\bf 16} 020127

\bibitem{Singh2008} Singh C 2008 Student understanding of quantum mechanics at the beginning of graduate instruction {\it Am. J. Phys.} {\bf 76} 277--87

\bibitem{Gonsar2021} Gonsar N, Patrick L and Cotner S 2021 Graduate- and undergraduate-student perceptions of and preferences for teaching practices in STEM classrooms {\it Discip. Interdiscip. Sci. Educ. Res.} {\bf 3} 6

\bibitem{GrossEOS2009} Gross N A \etal\ 2009 Space physics for graduate students: an activities-based approach {\it EOS Trans. AGU} {\bf 90} 13--4

\bibitem{LopezGross2008} Lopez R E and Gross N A 2008 Active learning for advanced students: the Center for Integrated Space Weather Modeling graduate summer school {\it Adv. Space Res.} {\bf 42} 1864--8

\bibitem{Lepage} Lepage G P 2021 Active learning in a graduate quantum field theory course {\it Am. J. Phys.} {\bf 89} 317--23

\bibitem{Deslauriers2019} Deslauriers L, McCarty L S, Miller K, Callaghan K and Kestin G 2019 Measuring actual learning versus feeling of learning in response to being actively engaged in the classroom {\it Proc. Natl Acad. Sci. USA} {\bf 116} 19251--7

\bibitem{Smith2009} Smith M K, Wood W B, Adams W K, Wieman C, Knight J K, Guild N and Su T T 2009 Why peer discussion improves student performance on in-class concept questions {\it Science} {\bf 323} 122--4

\bibitem{TurpenFinkelstein2009} Turpen C and Finkelstein N D 2009 Not all interactive engagement is the same: variations in physics professors' implementation of Peer Instruction {\it Phys. Rev. ST Phys. Educ. Res.} {\bf 5} 020101

\bibitem{LasryMazurWatkins2008} Lasry N, Mazur E and Watkins J 2008 Peer instruction: from Harvard to the two-year college {\it Am. J. Phys.} {\bf 76} 1066--9

\bibitem{Lasry2008} Lasry N 2008 Clickers or flashcards: is there really a difference? {\it Phys. Teach.} {\bf 46} 242--4

\bibitem{Henderson2012} Henderson C, Dancy M and Niewiadomska-Bugaj M 2012 Use of research-based instructional strategies in introductory physics: where do faculty leave the innovation-decision process? {\it Phys. Rev. ST Phys. Educ. Res.} {\bf 8} 020104

\bibitem{Dancy2024} Dancy M \etal\ 2024 Physics instructors' knowledge and use of active learning has increased over the last decade but most still lecture too much {\it Phys. Rev. Phys. Educ. Res.} {\bf 20} 010119

\bibitem{Wieman2013} Wieman C, Deslauriers L and Gilley B 2013 Use of research-based instructional strategies: how to avoid faculty quitting {\it Phys. Rev. ST Phys. Educ. Res.} {\bf 9} 023102
 
\bibitem{Mazur1997} Mazur E 1997 {\it Peer Instruction: A User's Manual} (Upper Saddle River, NJ: Prentice Hall)
 
\bibitem{Moldwin2023} Moldwin M B, Gross N and Cohen C 2023 Role and opportunity of solar and space physics summer schools to change culture {\it Bull. Am. Astron. Soc.} {\bf 55} 281

\bibitem{McCartyDeslauriers2020} McCarty L S and Deslauriers L 2020 Transforming a large university physics course to student-centered learning, without sacrificing content: a case study {\it The Routledge International Handbook of Student-Centered Learning and Teaching in Higher Education} ed S Hoidn and M Klemen\v{c}i\v{c} (Abingdon: Routledge) pp 186--200

\bibitem{Beatty2006} Beatty I D, Gerace W J, Leonard W J and Dufresne R J 2006 Designing effective questions for classroom response system teaching {\it Am. J. Phys.} {\bf 74} 31--9


\end{thebibliography}
\end{document}